\documentclass[pra,aps,twocolumn,superscriptaddress,notitlepage,10pt]{revtex4-1}

\usepackage[english]{babel}
\usepackage[T1]{fontenc}
\usepackage[utf8]{inputenc}
\usepackage{hyperref}
\usepackage{amsmath, amssymb, amsfonts,mathrsfs}
\usepackage{dsfont}
\usepackage{color}
\usepackage{graphicx}
\usepackage{times}
\usepackage{bm}

\DeclareMathOperator{\Id}{\mathds{1}}

\newcommand{\ea}{\textit{et al.}}

\newcommand{\bra}[1]{\left\langle{#1}\right\vert}
\newcommand{\ket}[1]{\left\vert{#1}\right\rangle}

\begin{document}

\title{Experimental detection of steerability in Bell local states with two measurement settings}

\author{Adeline Orieux}
\affiliation{LIP6, CNRS, Universit\'e Pierre et Marie Curie, Sorbonne Universit\'es, 75005 Paris, France}
\affiliation{IRIF, CNRS, Universit\'e Paris Diderot, Sorbonne Paris Cit\'e, 75013 Paris, France}

\author{Marc Kaplan}
\affiliation{LIP6, CNRS, Universit\'e Pierre et Marie Curie, Sorbonne Universit\'es, 75005 Paris, France}
\affiliation{School of Informatics, University of Edinburgh, Edinburgh, UK}

\author{Vivien Venuti}
\affiliation{LTCI, T\'el\'ecom ParisTech, Universit\'e Paris-Saclay, 75013 Paris, France}

\author{Tanumoy Pramanik}
\affiliation{Center for Quantum Information, Korea Institute of Science and Technology (KIST), Seoul, 02792, Republic of Korea}

\author{Isabelle Zaquine}
\affiliation{LTCI, T\'el\'ecom ParisTech, Universit\'e Paris-Saclay, 75013 Paris, France}

\author{Eleni Diamanti} 
\affiliation{LIP6, CNRS, Universit\'e Pierre et Marie Curie, Sorbonne Universit\'es, 75005 Paris, France}

\date{\today}

\begin{abstract}
Steering, a quantum property stronger than entanglement but weaker than non-locality in the quantum correlation hierarchy, is a key resource for one-sided device-independent quantum key distribution applications, in which only one of the communicating parties is trusted. A fine-grained steering inequality was introduced in [PRA 90 050305(R) (2014)], enabling for the first time the detection of steering in all steerable two-qubit Werner states using only two measurement settings. Here we numerically and experimentally investigate this inequality for generalized Werner states and successfully detect steerability in a wide range of two-photon polarization-entangled Bell local states generated by a parametric down-conversion source.
\end{abstract}

\maketitle


\section*{Introduction}

The notion of non-locality was first introduced in 1935 by Einstein, Podolsky and Rosen~\cite{EPR1935} by discussing a ``spooky action at a distance''. This led Schrödinger to introduce the concept of steering, in his response~\cite{ES1935,ES1936} to the EPR paper. The term steering describes a property of quantum mechanics that allows the subsystems of a bipartite entangled state to affect each other's state upon measurement: indeed the measurement of one of the subsystems can project (or steer) the other subsystem in a particular state, which depends on the chosen measurement on the first subsystem. In 2007, this concept was reformulated in terms of a quantum information task by Wiseman, Jones and Doherty~\cite{Wiseman2007}, which allowed them to establish a strict hierarchy between quantum correlations of increasing strength: entanglement, steering and Bell non-locality. Seen in this light, steering is the resource that allows Alice to win a game in which she tries to convince Bob (who does not trust her) that she can prepare an entangled state and share it between them.

As was shown in 2012 by Branciard~\ea~\cite{Branciard2012}, this resource can be exploited for quantum cryptography in a one-sided device-independent quantum-key-distribution (1SDI-QKD) scenario, intermediate between standard QKD in which both parties need to trust their measurement apparatus and device-independent QKD (DI-QKD)~\cite{Ekert1991,Acin2007} in which neither does: for 1SDI-QKD, only one of the parties needs a trusted measurement device. Compared to the DI-QKD scenario, this additional constraint of one trusted device must be fulfilled but, in return, the security can be based on the violation of a steering inequality rather than a Bell inequality, which lowers significantly the experimental requirements in terms of detection and transmission efficiencies~\cite{Zeilinger2012,Wiseman2012,White2012} because, in particular, the detection loophole needs to be closed only on Alice's side and the noise tolerance is higher.

In order to fully exploit this resource, practical and efficient ways of detecting steerability in experimentally-relevant bipartite states are needed. Over the last ten years, a lot of different steering inequalities have been proposed~\cite{Cavalcanti2009,Walborn2011,Schneeloch2013,Skrzypczyk2014,Cavalcanti2015,Kogias2015,Zhu2016} and experimentally tested~\cite{Pryde2010,White2012,Wiseman2012,Zeilinger2012,Pryde2015,Pryde2016,Guo2016} (for a recent review focused on semi-definite programming (SPD) see Ref.~\cite{SteeringReview2017}), however they are generally subject to some trade-off between the number of required measurement settings and their robustness to noise.
In particular, they require at least three measurement settings to detect steering in Bell-local Werner states~\cite{Pryde2010,Werner1989}.

In this article, we investigate numerically and experimentally a recently proposed steering inequality~\cite{Pramanik2014} based on fine-grained uncertainty relations~\cite{Wehner2010}, which is more efficient in that it allows the detection of steerability in a large range of noisy two-qubit states, in particular in all steerable Werner states, with two measurement settings. We first recall the fine-grained inequality in Section~\ref{sec:inequality}. In Section~\ref{sec:analysis} we apply the inequality to generalized Werner states, give a procedure to correctly choose the measurement settings and compare its performance with coarse-grained inequalities. Finally in Section~\ref{sec:experiment} we illustrate the procedure with a simple experiment.

\section{Fine-grained steering inequality}
\label{sec:inequality}

Let us first recall the fine-grained steering inequality introduced in Ref.~\cite{Pramanik2014} and the underlying game scenario. Alice prepares a bipartite state $\rho_{AB}$, keeps the part labeled $A$ for herself and sends the part labeled $B$ to Bob. Bob then asks Alice to steer $B$ to any eigenstate of an observable $U_B$, randomly chosen in the set $\{P,Q\}$ (with $P$ and $Q$ maximally incompatible). Alice measures $A$ with an observable $U_A$ (with $U_A=S$ if $U_B=P$ and $U_A=T$ if $U_B=Q$) and sends her measurement outcome $a\in\{0;1\}$ to Bob. Bob finally measures $B$ with the previously chosen observable $U_B$ and gets an outcome $b\in\{0;1\}$. After repeating these steps a large number of times, Bob is convinced that Alice can indeed steer his subsystem (i.e. that $\rho_{AB}$ is steerable) if the following inequality is violated:
\begin{eqnarray}
&F^{a,b}& = \frac{1}{2} \Big(\textrm{P}(b_P|a_S)+\textrm{P}(b_Q|a_T)\Big) \nonumber \\
&\,&\leq F_{\textrm{lim}} = \frac{1}{2} \displaystyle \max_{P^*,Q^*} \Big[ \displaystyle \max_{\lambda} \big[\textrm{P}_q(b_{P^*}|\lambda)+\textrm{P}_q(b_{Q^*}|\lambda)\big]\Big]
\label{EqSteering}
\end{eqnarray}
where $\textrm{P}(b_{P,Q}|a_{S,T})$ is the conditional probability of Bob getting the outcome $b$ upon measurement of $P$ or $Q$ when Alice announces the outcome $a$, $P^*$ and $Q^*$ range over all possible maximally incompatible measurements, and $\textrm{P}_q(b|\lambda)$ is the probability of obtaining an outcome $b$ upon a quantum measurement on a quantum system $\rho_B$ described by a hidden variable $\lambda$. The left-hand side $F^{a,b}$ of the inequality is the steering parameter which can be estimated from measurement results. The right-hand side $F_{\textrm{lim}}$ is calculated by considering a local hidden state (LHS) model for $\rho_B$: this model corresponds to the cheating strategy of a dishonest Alice who prepares a local state $\rho_B$ instead of an entangled state $\rho_{AB}$ and announces values of $a$ that do not correspond to actual quantum measurements. As was shown in Ref.~\cite{Pramanik2014}, if Alice knows beforehand the set $\{P,Q\}$ of Bob's possible measurement settings (``scenario I''), then $F_{\textrm{lim}}(\textrm{I})=(1+1/\sqrt{2})/2 \approx 0.854$. If, however, Alice does not know the set $\{P,Q\}$ before preparing $\rho_{AB}$ (``scenario II''), then $F_{\textrm{lim}}(\textrm{II})=3/4$.

\begin{figure}[t]
\centering
\includegraphics[width=0.40\textwidth]{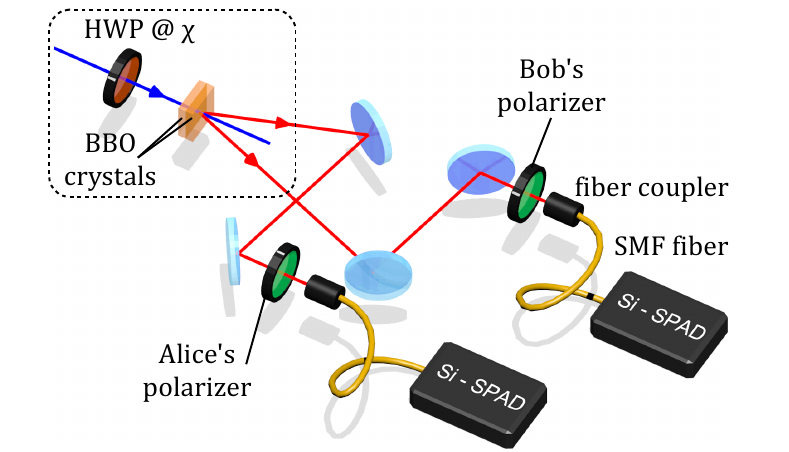}
\caption{(Color online) Experimental set-up (for more details see~\cite{QuTools}). The main elements of the source are shown in the dashed-line box: a CW laser beam at 405~nm pumps two BBO crystals with orthogonal axes. The angle $\chi$ of the half-wave plate (HWP) controls the amount of photon pairs generated in the first and second crystals. At Alice's and Bob's station, projective measurements are implemented with a rotating polarizer. The photons at 810~nm are coupled into single-mode fibers and detected in coincidence by silicon avalanche photodiodes (Si-SPAD) and timing electronics. Ceiling lamps (not shown) are used to fix the amount of unpolarized background noise.}
\label{Fig:Setup}
\end{figure}

\section{Numerical analysis for generalized Werner states}
\label{sec:analysis}

In the following, we study the aforementioned inequality in both scenarios for states of the form:
\begin{equation}
\rho = p\left(\eta\ket{\Phi^+_\alpha}\bra{\Phi^+_\alpha}+(1-\eta)\ket{\Phi^-_\alpha}\bra{\Phi^-_\alpha}\right) + \frac{(1-p)}{4}\Id_A\otimes\Id_B,
\label{EqSPDCstate}
\end{equation}
where $\ket{\Phi^\pm_\alpha}=\sqrt{\alpha}\ket{00}_{AB}\pm e^{i\varphi}\sqrt{1-\alpha}\ket{11}_{AB}$ is a pure state, $\alpha, p,\eta \in [0;1]$, $\Id$ is the $2\times2$ identity matrix and subscripts $A,B$ refer to Alice's and Bob's qubit respectively. These states model well polarization-entangled two-photon pairs produced e.g. by bulk type-I spontaneous parametric down-conversion sources~\cite{Kwiat1999} (see the dashed-line box in Fig.~\ref{Fig:Setup}) where a laser beam, with a linear polarization adjusted by a half-wave plate (HWP) set at an angle $\chi$, pumps two adjacent nonlinear crystals having optical axes orthogonal to each other. Then $\alpha = \cos^2(2\chi)$ corresponds to the amount of pairs emitted in the first crystal (horizontally-polarized photons, denoted by $\ket{00}$) with respect to those emitted in the second crystal (vertically-polarized photons, denoted by $\ket{11}$); $\varphi$ is the phase between the two possible pair emission processes; $p$ is linked to the amount of unpolarized noise present in the set-up (background light, fluorescence of the optical elements...); and $\eta$ is linked to a dephasing noise accounting for the partial distinguishability between the optical modes of photons emitted by each crystal.

\begin{figure}[b]
\centering
\includegraphics[width=0.45\textwidth]{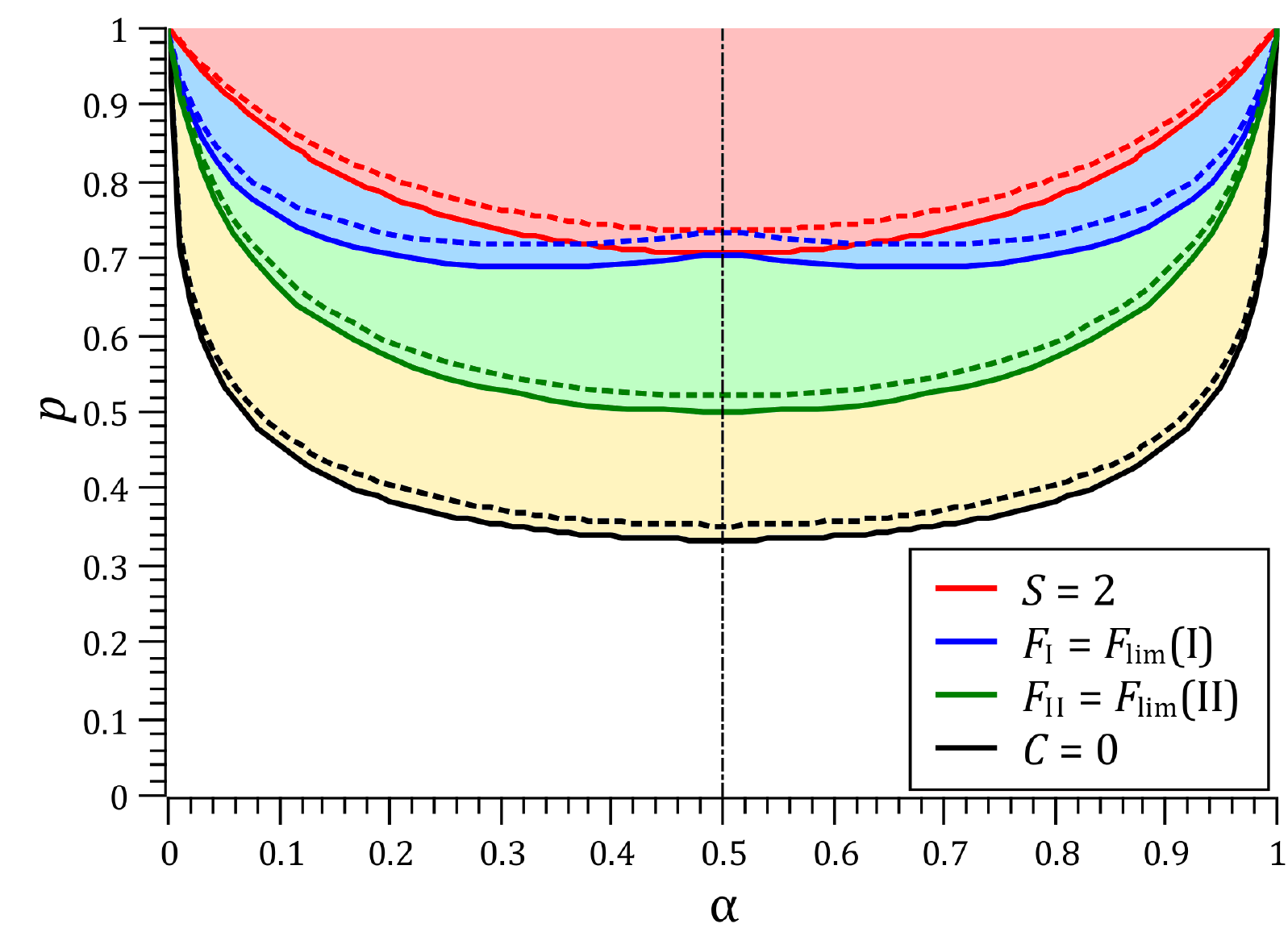}
\caption{(Color online) Lower bounds on $p$ as a function of $\alpha$ for $\eta=1$ (full lines) and $\eta=0.96$ (dotted lines) for Bell non-locality (CHSH: red lines), steering (scenario I: blue lines, scenario II: green lines) and entanglement (black lines). Shaded red area: Bell non-local states; green and blue areas: steerable states; yellow area: entangled states. Vertical mixed black line: Werner state, $\alpha = 1/2$.}
\label{Fig:pbounds}
\end{figure}

For such states, one can show (see Appendix~\ref{AppendixA}) that the concurrence~\cite{Wootters1998} (a tight entanglement witness with $C>0$ if and only if the state is entangled) is given by~\cite{Yu2007}:
\begin{equation}
C = 2\textrm{max}\left( 0 , p(2\eta-1)\sqrt{\alpha(1-\alpha)}-\frac{1-p}{4} \right),
\label{Eq:RhoC}
\end{equation}
and that the maximal Bell parameter of the CHSH inequality~\cite{CHSH,Gisin1991} ($S\leq2$ for local hidden variable (LHV) models and $2<S\leq2\sqrt{2}$ for Bell non-local quantum states) is:
\begin{equation}
S = 2p\sqrt{1+4(2\eta-1)^2\alpha(1-\alpha)}.
\label{Eq:RhoS}
\end{equation}
These expressions, for given values of $\eta$ and $\alpha$, give lower bounds on the value of $p$ for which $\rho$ is entangled or Bell non-local (see black and red lines in Fig.~\ref{Fig:pbounds}). In particular, for Werner states~\cite{Werner1989} (i.e. $\eta=1$ and $\alpha=1/2$), the state is entangled if and only if $p>1/3$ and it is Bell non-local if $p>1/\sqrt{2}$.
Note that here we use the CHSH inequality to distinguish between Bell non-local states and states admitting a LHV model (which we call Bell-local here) since we restrict Alice and Bob to only two projective measurement settings~\cite{Acin2006}. This would not be valid for an arbitrary number of projective measurements since, in that case, it has been shown~\cite{Acin2006} that, for Werner states, the separation occurs for $p=1/K_G(3)$, $K_G(3)$ being Grothendiek's constant of order 3 for which the best known lower~\cite{Vertesi2016} and upper~\cite{Hirsch2017} bounds up to now are $1.4261 \leq K_G(3) \leq 1.4644$, i.e. $0.7012 \geq p \geq 0.6829$. Hence, strictly speaking, we are sure that a Werner state is Bell-local only if $p<0.6829$.

To compute the steering parameter defined in Eq.~(\ref{EqSteering}), any combination $\{a,b\}$ can be chosen with $a,b \in \{0;1\}$ but some choices may give larger values than others depending on the asymmetry of the state, thus we will define a more general steering parameter $F = \max_{a,b}\left(F^{a,b}\right)$. Writing Bob's and Alice's measurement settings as $U = \cos\theta_U\sigma_z + \sin\theta_U\big(\cos\phi_U\sigma_x+\sin\phi_U\sigma_y\big)$ (with $U=P,Q,S,T$ and $\sigma_x$, $\sigma_y$ and $\sigma_z$ the Pauli matrices), and imposing $\theta_Q=\theta_P+\pi/2$ and $\phi_Q=\phi_P$ such that $P$ and $Q$ are maximally incompatible measurements, $F$ can be expressed as:
\begin{eqnarray}
&F& = \frac{1}{4}\Big[\big[ 1 + p\big(\xi \mathcal{C}_P+(\xi+\mathcal{C}_P)\mathcal{C}_S)+\zeta_S\mathcal{S}_P\mathcal{S}_S\big) \big]/\big[1+p\xi \mathcal{C}_S\big] \nonumber\\
&+& \big[ 1 + p\big(-\xi \mathcal{S}_P+(\xi-\mathcal{S}_P)\mathcal{C}_T)+\zeta_T\mathcal{C}_P\mathcal{S}_T\big) \big]/\big[1+p\xi \mathcal{C}_T\big]\Big] \nonumber\\
\label{Eq:RhoSteeringPr}
\end{eqnarray}
where $\xi=\sigma|2\alpha-1|$, $\zeta_A=2(2\eta-1)\sqrt{\alpha(1-\alpha)}\cos(\varphi-\phi_A)$ ($A=S,T$), $\mathcal{C}_U=\cos\theta_U$ and $\mathcal{S}_U=\sin\theta_U$ ($U=P,S,T$), and $\sigma$ denotes the sign of $\cos(\theta_P+\pi/4)$ (see Appendix~\ref{AppendixA} for details).
The angles $\theta_S$, $\phi_S$, $\theta_T$ and $\phi_T$ are optimized so as to maximize $F$ once $P$ and $Q$ have been fixed.
For scenario I, Bob's measurements $P$ and $Q$ are fixed to $\sigma_z$ and $\sigma_x$ respectively, thus $F_{\textrm{I}}= \max_{S,T}(F)$ with $\theta_P=\phi_P=\phi_Q=0$ and $\theta_Q=\pi/2$.
For scenario II, $P$ and $Q$ must be chosen such that $F$ is minimized: $F_{\textrm{II}}= \min_{P,Q}(\max_{S,T}(F))$. Note that for Werner states (i.e. $\alpha=1/2$ and $\eta=1$) which are symmetric states, any choice of maximally incompatible settings $P$ and $Q$ will give the same value of $F$; however for $\alpha \neq 1/2$ this is not the case and the minimization must be done in order to avoid overestimating $F$ (see Appendix~\ref{AppendixB}). This constraint comes from the requirements that $P$ and $Q$ should be unknown by Alice when she prepares $\rho_{AB}$.

\begin{figure}[ht]
\centering
\includegraphics[width=0.45\textwidth]{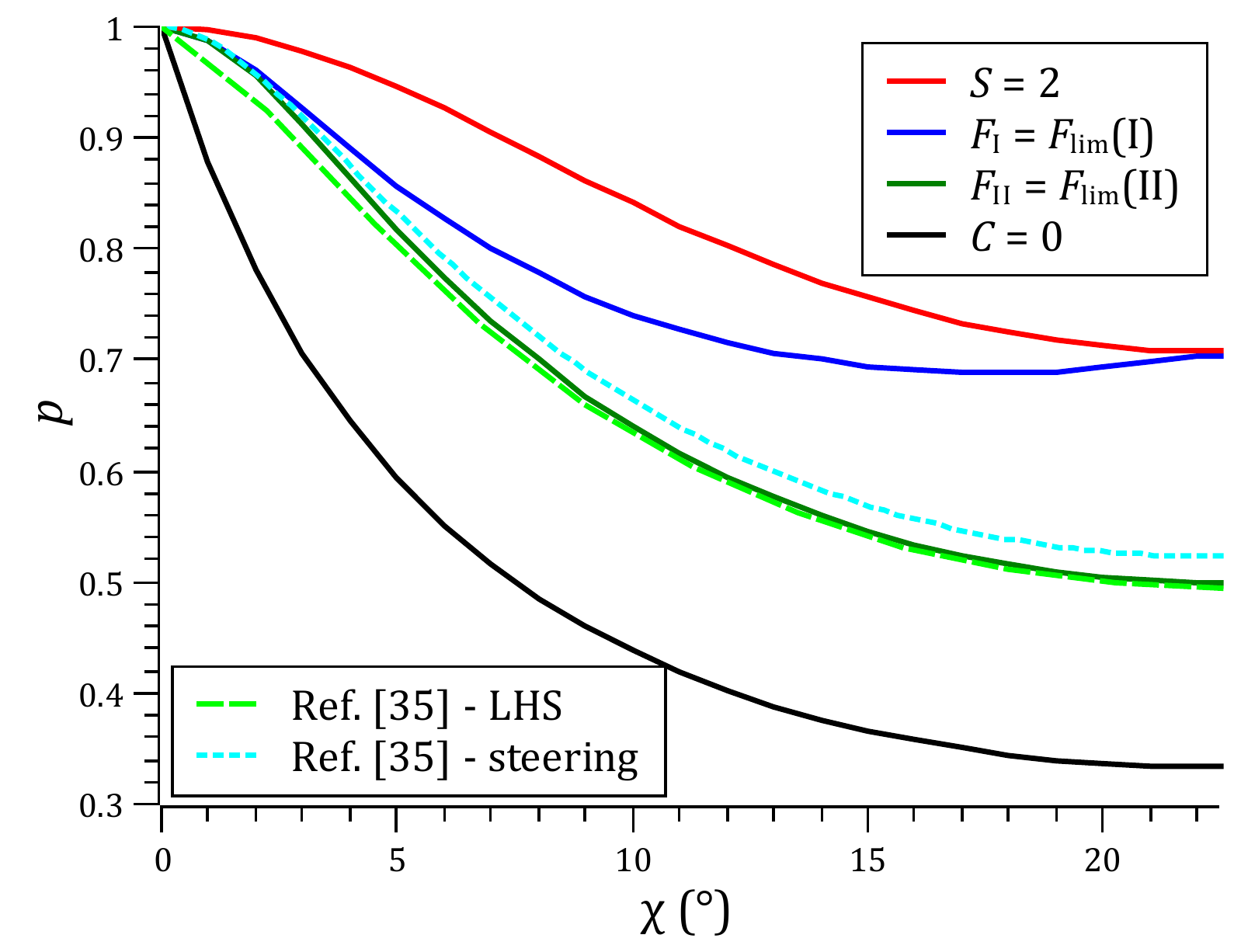}
\caption{Bounds on $p$ as a function of $\chi$ (in degrees), for a generalized Werner state with $\eta=1$. The thick red, dark blue, dark green and black lines are the lower bounds presented in Fig.~\ref{Fig:pbounds}. The light green dashed line is an upper bound for non-steerable states found in Ref.~\cite{Brunner2016} by a numerical iterative construction of a LHS model with 4 steps. The light blue dotted line is a lower bound for steerable states found also in Ref.~\cite{Brunner2016} with the SDP method of Ref.~\cite{Skrzypczyk2014} and 9 measurement settings. The steering lower bound given by the fine-grained steering inequality in scenario II (dark green line) lies in between these two numerical bounds.}
\label{Fig:BoundII}
\end{figure}

Equation~(\ref{Eq:RhoSteeringPr}) gives a lower bound on $p$ for the state $\rho$ to be steerable, according to scenario I or II (see blue and green lines in Fig.~\ref{Fig:pbounds}). For the particular case of Werner states, using scenario I, we find a lower bound of $1/\sqrt{2}$ for $p$, the same as for Bell non-locality detected by the CHSH inequality and the same as for all coarse-grained steering inequalities with two measurement settings~\cite{SteeringReview2017}. However, using scenario II, the theoretical bound $p=1/2$~\cite{Wiseman2007} is reached with only two measurement settings.
Note that we can even conjecture that scenario II gives an optimal lower bound for generalized Werner states (i.e. for any $\alpha\in]0;1[$). Indeed Fig.~\ref{Fig:BoundII} shows that this bound (dark green full line) lies between a lower bound for steerable states (light blue dotted line) and an upper bound for states with a LHS model (light green dashed line) that were both calculated numerically in Ref.~\cite{Brunner2016}, respectively with a semi-definite program~\cite{Skrzypczyk2014} and with a new iterative method for constructing LHS models~\cite{Brunner2016,Cavalcanti2016}.
We can also notice that, for both scenarios, the set of steerable states detected with two settings is strictly larger than the set of Bell non-local states seen by CHSH: for $\alpha \in ]0;\frac{1}{2}[\, \cup \, ]\frac{1}{2};1[$, the lower bound on $p$ for steering is strictly lower than the one for Bell non-locality. In particular, even for scenario I, Bell local steerable states can be detected for a large range of values of $\alpha$.

\begin{figure*}[t]
\centering
\includegraphics[width=1\textwidth]{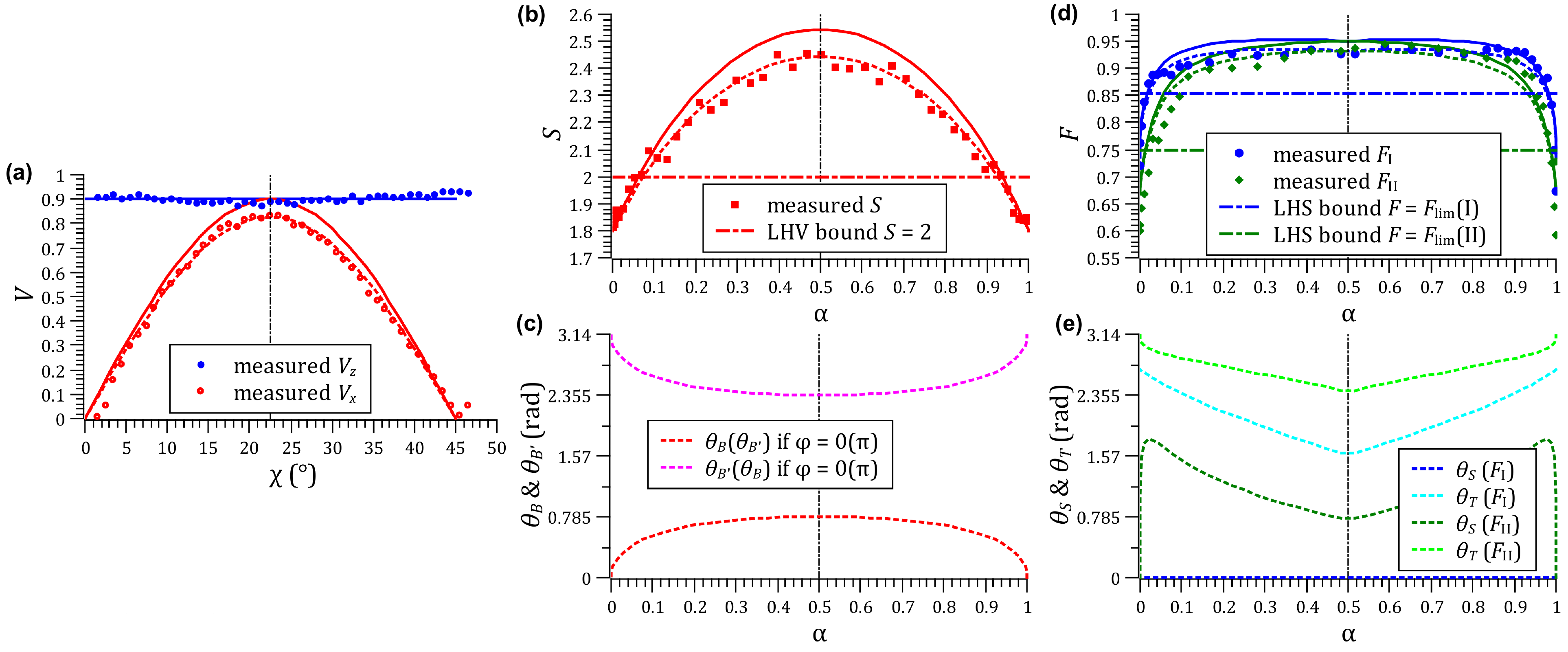}
\caption{(Color online) (a) Measured visibilities in the $\sigma_z$ (blue dots) and $\sigma_x$ (red open circles) bases as a function of the angle $\chi$ of the pump's half-wave plate. Error bars (not shown) are about the size of the symbols or smaller. Solid lines: simulations for $p=0.90$ and $\eta=1$. Dotted lines: simulations for $p=0.90$ and $\eta=0.96$. (b) Bell and (d) steering parameters as a function of $\alpha$. Symbols: measurements, full lines: simulations for $p=0.90$ and $\eta=1$, dotted lines: simulations for $p=0.90$ and $\eta=0.96$. Bell parameter $S$: red squares and lines, steering parameters $F_{\textrm{I}}$: blue circles and lines, and $F_{\textrm{II}}$: green diamonds and lines. The mixed horizontal lines show the LHV limit $S_{\textrm{lim}}=2$ and the LHS limits $F_{\textrm{lim}}(\textrm{I}) =(1+1/\sqrt{2})/2$ and $F_{\textrm{lim}}(\textrm{II}) = 3/4$. Vertical error bars (not shown) stemming from Poisson photon counting statistics are $<0.025$ for $S$ (about the symbol size) and $<0.005$ ($\alpha \in [0.1;0.9]$) or $<0.02$ ($\alpha<0.1$ or $\alpha>0.9$) for $F$ (see Appendix~\ref{AppendixC}). (c) optimized CHSH measurement settings (see Appendix~\ref{AppendixA}) $B,B'=\cos(\theta_{B,B'})\sigma_z+\sin(\theta_{B,B'})\sigma_x$ for Bob when fixing Alice's measurements to $A=\sigma_z$ ($\theta_A = 0$) and $A'=\sigma_x$ ($\theta_A' = \pi/2$). (e) optimized steering measurement settings for Alice for scenario I ($\theta_S$ in blue and $\theta_T$ in light blue) and for scenario II ($\theta_S$ in green and $\theta_T$ in light green), calculated with \textit{Wolfram Alpha}~\cite{Wolfram} for $p=0.90$ and $\eta=0.96$ (see Appendix~\ref{AppendixA}). In all plots, the vertical mixed line corresponds to $\alpha=1/2$.}
\label{Fig:results}
\end{figure*}

\section{Experimental implementation}
\label{sec:experiment}

We have experimentally tested this fine-grained steering inequality in both scenarios with a commercial source of polarization-entangled two-photon states (QuTools~\cite{QuTools}) based on the scheme of Ref.~\cite{Kwiat1999} (see Fig.~\ref{Fig:Setup}). The projective measurements corresponding to settings $U_A$ on Alice's photon and $U_B$ on Bob's photon, with $U_{A,B}=\cos(\theta_{A,B})\sigma_z + \sin(\theta_{A,B})\sigma_x$ (i.e. in the $(\sigma_z;\sigma_x)$ plane of Bloch's sphere), are implemented with a polarizer whose axis is set at an angle $\theta_{A,B}/2$ (projection on the $+1$ eigenstate of $U_{A,B}$) or $\theta_{A,B}/2+\pi/2$ (projection on the $-1$ eigenstate of $U_{A,B}$) with respect to the vertical direction. Note that $\sigma_y$ measurements are not needed when $\varphi=0$ or $\pi$ (see Appendix~\ref{AppendixA}).

We first characterized the experimental state with visibility measurements $V_z$ ($V_x$) in the $\sigma_z$ ($\sigma_x$) basis, for different values of $\chi$ (Fig.~\ref{Fig:results}(a)). Modeling the state with Eq.~(\ref{EqSPDCstate}) and using the relation $\alpha = \cos^2(2\chi)$, one can show that $V_z = p$ and $V_x = 2p(2\eta-1)\sqrt{\alpha(1-\alpha)}$ (see Appendix~\ref{AppendixA}). We thus deduced a noise parameter $p = 0.90$, a dephasing parameter $\eta = 0.96$ and a phase $\varphi = \pi$.

In Fig.~\ref{Fig:results}(b) we show the measured value of the Bell parameter $S=\textrm{E}(A,B)-\textrm{E}(A',B)+\textrm{E}(A,B')+\textrm{E}(A',B')$, with $\textrm{E}(A,B)=\textrm{P}(0_A,0_B)+\textrm{P}(1_A,1_B)-\textrm{P}(1_A,0_B)-\textrm{P}(0_A,1_B)$, and its theoretical value given by Eq.~(\ref{Eq:RhoS}), for the optimized measurement settings $A,A',B,B'$~\cite{Gisin1991} shown in Fig.~\ref{Fig:results}(c). The state violates the CHSH inequality $S\leq2$ and is thus non-local for $\alpha \in ]0.075;0.925[$, with a maximal value of $S=2.45$ for $\alpha=1/2$.
The corresponding measured values of the steering parameter in both scenarios $F_\textrm{I}$ and $F_\textrm{II}$ are shown in Fig.~\ref{Fig:results}(d). For scenario II, Bob's measurement angles have been set to $\phi_P=\phi_Q=0$ and $\theta_P=\pi/4$, $\theta_Q=3\pi/4$ which minimize Eq.~(\ref{Eq:RhoSteeringPr}) for $F$ (for any value of $\alpha$, $p$ and $\eta$, see Appendix~\ref{AppendixB}). For both scenarios, Alice's settings have been optimized (with $\phi_S=\phi_T=\varphi=\pi$) for each value of $\alpha$ so as to maximize $F$ and are shown in Fig.~\ref{Fig:results}(e). The state violates the fine-grained steering inequality (Eq.~(\ref{EqSteering})) in scenario I for $\alpha \in ]0.022;0.978[$ and in scenario II for $\alpha \in ]0.015;0.985[$. For scenario I, the maximum value of $F_\textrm{I}=0.935$ is obtained for $\alpha=0.35$ and $\alpha=0.65$. For scenario II, the maximum is $F_\textrm{II}=0.932$ for $\alpha=1/2$.

\begin{figure}[t]
\centering
\includegraphics[width=0.5\textwidth]{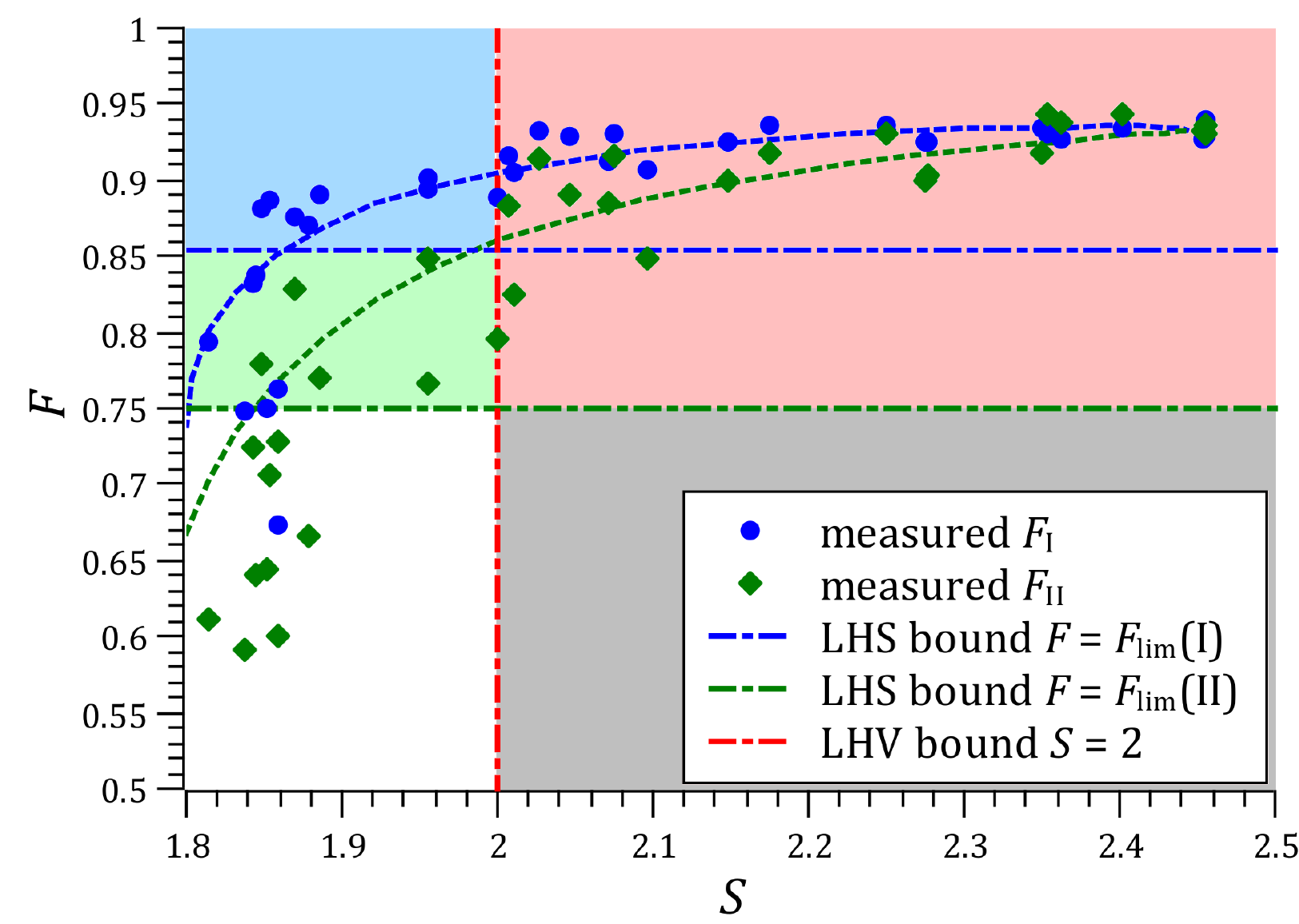}
\caption{(Color online) Steering parameter $F$ as a function of the Bell parameter $S$. Symbols: measurements, dotted lines: simulations for $p=0.90$ and $\eta=0.96$. $F_I$: blue circles and lines, and $F_{II}$: green diamonds and lines. The mixed lines show the LHS and LHV limits.}
\label{Fig:resultsFvsS}
\end{figure}

In Fig.~\ref{Fig:resultsFvsS}, we have plotted the measured values of $F_\textrm{I}$ and $F_\textrm{II}$ against $S$, together with the simulations for $p=0.90$ and $\eta=0.96$. We can identify five main zones in this plot, corresponding to quantum correlations of different strength. The shaded red area corresponds to Bell non-local states that violate the CHSH inequality (i.e. $S>2$) and also violate the fine-grained steering inequality. The blue (green) area corresponds to steerable Bell local states detected in scenario I or II: $F_\textrm{I} > (1+1/\sqrt{2})/2$ or $F_\textrm{II} > 3/4$, and $S \leq 2$. The white area corresponds to states that violate neither the CHSH inequality nor the fine-grained steering inequality. Finally the grey area corresponds to states that are Bell non-local but unsteerable which is in contradiction with the established hierarchy of quantum correlations~\cite{Wiseman2007}; none of our measurement results fall in this zone.

\section{Conclusion}

In conclusion, we have numerically and experimentally investigated the performance of the fine-grained steering inequality of Ref.~\cite{Pramanik2014} for the detection of steerability in experimentally-relevant two-photon states which can be modeled as generalized Werner states with dephasing. We have shown that contrary to other steering inequalities which are coarse-grained~\cite{SteeringReview2017} and require strictly more than two measurement settings to detect steerability in Bell local states~\cite{Pryde2010}, this fine-grained inequality is able to detect a much larger set of steerable states with only two measurement settings, in particular all steerable Werner states and (most probably) all steerable generalized Werner states. We have also shown that even using the most conservative LHS bound of scenario I, the inequality allows the detection of steerability in states that do not violate the Bell-CHSH inequality. Finally, for scenario I, a key rate $r \geq \log_2[F_\textrm{I}/(2F_{\textrm{lim}}(\textrm{I})-F_\textrm{I})]$ for 1SDI-QKD has been proven in Ref.~\cite{Pramanik2014} against individual eavesdropping attacks; with our best value of $F_\textrm{I}=0.935$, a rate $r \geq 0.276$ secure bit per photon detected by Bob could thus be achieved. It will be interesting to extend this calculation to scenario II, which allows for a better noise tolerance.

\section*{Acknowledgements}
We thank Ioannis Touloupas for his help with the data acquisition, and Nicolas Brunner and Flavien Hirsch for kindly providing numerical data for steering and LHS bounds in Ref.~\cite{Brunner2016}. We acknowledge financial support from the BPI France project RISQ, the French National Research Agency projects QRYPTOS and COMB, and the Ile-de-France
Region project QUIN. The work of MK is supported in part by EPSRC grant number EP1N003829/1 Verification of Quantum Technology.

\appendix

\section{Detailed analytical calculations for generalized Werner states and optimal measurement angles}
\label{AppendixA}

\subsection{Projective measurement outcomes}

When applying to the generalized Werner state $\rho$ defined in Eq.~(\ref{EqSPDCstate}) a projective measurement described by an observable (i.e. a measurement setting) $A \otimes B$, with
\begin{equation}
U = \cos\theta_U\sigma_z + \sin\theta_U\big(\cos\phi_U\sigma_x+\sin\phi_U\sigma_y\big),
\label{EqU}
\end{equation}
where $\sigma_x$, $\sigma_y$ and $\sigma_z$ are the Pauli matrices and $U=A,B$, the computational basis states are transformed as:
\begin{eqnarray}
\ket{0}_U &\rightarrow& \cos\frac{\theta_U}{2}\ket{0}_U+\sin\frac{\theta_U}{2}\textrm{e}^{i\phi_U}\ket{1}_U, \nonumber \\
\ket{1}_U &\rightarrow& -\sin\frac{\theta_U}{2}\ket{0}_U+\cos\frac{\theta_U}{2}\textrm{e}^{i\phi_U}\ket{1}_U, \nonumber
\label{EqP00}
\end{eqnarray}
and the probabilities of the four possible combinations of outcomes $a$ for $A$ and $b$ for $B$ are:
\begin{equation}
\textrm{P}(a_A,a_B) = \langle ab |(A\otimes B)\rho (A\otimes B)^\dagger| ab \rangle, \nonumber
\label{EqPab}
\end{equation}
which gives:
\begin{eqnarray}
\textrm{P}(0_A,0_B) &=& \frac{1}{4} + \frac{p}{4}\Big[ (2\alpha-1)\cos\theta_B + \big(2\alpha-1+\cos\theta_B\big)\cos\theta_A \nonumber \\
&+& (2\eta-1)\sqrt{\alpha(1-\alpha)}\cos(\varphi-\phi_A-\phi_B)\sin\theta_B\sin\theta_A \Big],\nonumber
\label{EqP00}
\end{eqnarray}
\begin{eqnarray}
\textrm{P}(0_A,1_B) &=& \frac{1}{4} + \frac{p}{4}\Big[ (1-2\alpha)\cos\theta_B - \big(1-2\alpha+\cos\theta_B\big)\cos\theta_A \nonumber \\
&-& (2\eta-1)\sqrt{\alpha(1-\alpha)}\cos(\varphi-\phi_A-\phi_B)\sin\theta_B\sin\theta_A \Big],\nonumber
\label{EqP01}
\end{eqnarray}
\begin{eqnarray}
\textrm{P}(1_A,0_B) &=& \frac{1}{4} + \frac{p}{4}\Big[ (2\alpha-1)\cos\theta_B - \big(2\alpha-1+\cos\theta_B\big)\cos\theta_A \nonumber \\
&-& (2\eta-1)\sqrt{\alpha(1-\alpha)}\cos(\varphi-\phi_A-\phi_B)\sin\theta_B\sin\theta_A \Big],\nonumber
\label{EqP10}
\end{eqnarray}
\begin{eqnarray}
\textrm{P}(1_A,1_B) &=& \frac{1}{4} + \frac{p}{4}\Big[ (1-2\alpha)\cos\theta_B + \big(1-2\alpha+\cos\theta_B\big)\cos\theta_A \nonumber \\
&+& (2\eta-1)\sqrt{\alpha(1-\alpha)}\cos(\varphi-\phi_A-\phi_B)\sin\theta_B\sin\theta_A \Big],\nonumber
\label{EqP11}
\end{eqnarray}
The correlation function of this measurement is
\begin{eqnarray}
\textrm{E}(A \otimes B) &=& \textrm{P}(a_A = b_B) - \textrm{P}(a_A = \bar{b}_B) \nonumber \\
&=& \textrm{P}(0_A,0_B)+\textrm{P}(1_A,1_B)-\textrm{P}(0_A,1_B)-\textrm{P}(1_A,0_B). \nonumber
\label{EqE}
\end{eqnarray}

\subsection{Visibilities $V_z$ and $V_x$}

The visibility is $V_z = |\textrm{E}(\sigma_z \otimes \sigma_z)| = p$ in the $\sigma_z$ basis and $V_x = |\textrm{E}(\sigma_x \otimes \sigma_x)| = 2p(2\eta-1)\sqrt{\alpha(1-\alpha)}$ in the $\sigma_x$ basis.

\subsection{Concurrence $C$}

For an X-shaped density matrix of the form
\begin{equation*}
 \begin{pmatrix}
  a & 0 & 0 & w \\
  0 & b & z & 0 \\
  0 & z^* & c & 0 \\
  w^* & 0 & 0 & d
 \end{pmatrix},
\end{equation*}
Wootters' concurrence~\cite{Wootters1998} can be simply calculated~\cite{Yu2007} as $C = 2\max\left( 0 , |z|-\sqrt{ad} , |w|-\sqrt{bc}\right)$, which for the state of Eq.~(\ref{EqSPDCstate}) gives Eq.~(\ref{Eq:RhoC}):
\begin{equation}
C = 2\max\left(0 , p(2\eta-1)\sqrt{\alpha(1-\alpha)}-\frac{1-p}{4}\right). \nonumber
\label{EqC}
\end{equation}

\subsection{Bell-CHSH parameter $S$}
The Bell parameter for the CHSH inequality $S \leq 2$~\cite{CHSH} is given by:
\begin{eqnarray}
S &=& |\textrm{E}(A \otimes B)-\textrm{E}(A \otimes B')+\textrm{E}(A' \otimes B)+\textrm{E}(A' \otimes B')| \nonumber \\
&=& p\Big( \cos\theta_{A}(\cos\theta_{B} -\cos\theta_{B'}) + \cos\theta_{A'}(\cos\theta_{B} +\cos\theta_{B'}) \nonumber \\
&\,& + 2(2\eta-1)\sqrt{\alpha(1-\alpha)}\cos\varphi \nonumber \\
&\,& \times \big( \sin\theta_{A}(\sin\theta_{B} -\sin\theta_{B'}) +\sin\theta_{A'}(\sin\theta_{B} +\sin\theta_{B'}) \big) \Big), \nonumber
\label{EqS}
\end{eqnarray}
where $A,A',B,B'$ are four measurement settings described by Eq.~(\ref{EqU}) with $\phi_A=\phi_{A'}=\phi_B=\phi_{B'}=0$ (although one could use different values of $\phi$ in case $\varphi \neq 0,\pi$).\\
The optimal measurement angles do not depend on the value of $p$ and can be calculated as in Ref.~\cite{Gisin1991}: fixing $\theta_A=0$ ($A=\sigma_z$) and $\theta_{A'}=\pi/2$ ($A'=\sigma_x$), the optimal measurement angles for Bob (that maximize $S$) are:
\begin{equation}
\theta_{B,B'} = 2\,\textrm{atan}\left(\frac{\sqrt{4(2\eta-1)^2\alpha(1-\alpha)+1} \pm 1}{2(2\eta-1)\sqrt{\alpha(1-\alpha)}}\right),
\label{EqSthetaB}
\end{equation}
and the corresponding maximal value of the Bell parameter (Eq.~(\ref{Eq:RhoS})) is:
\begin{equation}
S = 2p\sqrt{1+4(2\eta-1)^2\alpha(1-\alpha)}. \nonumber
\label{EqSmax}
\end{equation}

\subsection{Fine-grained steering parameter $F$}
The fine-grained steering parameter given in Eq.~(\ref{Eq:RhoSteeringPr}) of the main text is calculated as follows:
\begin{equation}
F = \displaystyle \max_{a,b}\left(F^{a,b}\right) = \max \left(F^{0,0},F^{1,1}\right), \nonumber
\label{EqF}
\end{equation}
with $F^{a,b} = \textrm{P}(b_P|a_S) + \textrm{P}(b_Q|a_T)$ and $\textrm{P}(b_B|a_A) = \textrm{P}(a_A,b_B)/\left(\textrm{P}(a_A,b_B)+\textrm{P}(a_A,\bar{b}_B)\right)$, with $A=S,T$ and $B=P,Q$.\\
Thus, if we impose that $\theta_Q=\theta_P+\pi/2$ and $\phi_Q=\phi_P=0$ (one possible choice for two maximally incompatible measurement operators $P$ and $Q$ for Bob), we have, for any choice of $\theta_P$:
\begin{eqnarray}
F^{0,0} &=& \Big[ 1 + p\big[ (2\alpha-1)\cos\theta_P + \big(2\alpha-1+\cos\theta_P\big)\cos\theta_S \nonumber\\
&\,& + 2(2\eta-1)\sqrt{\alpha(1-\alpha)}\cos(\varphi-\phi_S)\sin\theta_P\sin\theta_S \big] \Big] \nonumber\\
&\,& \times \frac{1}{4\big[1+p(2\alpha-1)\cos\theta_S\big]} \nonumber\\
&+& \Big[ 1 + p\big[ -(2\alpha-1)\sin\theta_P + \big(2\alpha-1-\sin\theta_P\big)\cos\theta_T \nonumber\\
&\,& + 2(2\eta-1)\sqrt{\alpha(1-\alpha)}\cos(\varphi-\phi_T)\cos\theta_P\sin\theta_T \big] \Big] \nonumber\\
&\,& \times \frac{1}{4\big[1+p(2\alpha-1)\cos\theta_T\big]}, \nonumber
\end{eqnarray}
and
\begin{eqnarray}
F^{1,1} &=& \Big[ 1 + p\big[ (1-2\alpha)\cos\theta_P + \big(1-2\alpha+\cos\theta_P\big)\cos\theta_S \nonumber\\
&\,& + 2(2\eta-1)\sqrt{\alpha(1-\alpha)}\cos(\varphi-\phi_S)\sin\theta_P\sin\theta_S \big] \Big] \nonumber\\
&\,& \times \frac{1}{4\big[1+p(1-2\alpha)\cos\theta_S\big]} \nonumber \\
&+& \Big[ 1 + p\big[ -(1-2\alpha)\sin\theta_P + \big(1-2\alpha-\sin\theta_P\big)\cos\theta_T \nonumber\\
&\,& + 2(2\eta-1)\sqrt{\alpha(1-\alpha)}\cos(\varphi-\phi_T)\cos\theta_P\sin\theta_T \big] \Big] \nonumber \\
&\,& \times \frac{1}{4\big[1+p(1-2\alpha)\cos\theta_T\big]}. \nonumber
\end{eqnarray}
$F^{0,0}$ is maximized for

$\left \{
\begin{array}{c @{\,=\,} c}
    X\sin\theta_P\left[\cos\theta_S+p(2\alpha-1)\right]+Y\cos\theta_P\sin\theta_S & 0 \\
    X\cos\theta_P\left[\cos\theta_T+p(2\alpha-1)\right]-Y\sin\theta_P\sin\theta_T & 0, \\
\end{array}
\right.$
which gives, for $\theta_P \neq 0,\pi$:
\begin{eqnarray}
\theta_S = 2\,&\textrm{atan}&\Big[\frac{1}{X\sin\theta_P\left[1-p(2\alpha-1)\right]} \times\Big(Y\cos\theta_P \nonumber \\
&\,& +\sqrt{X^2\sin^2\theta_P\left[1-p^2(2\alpha-1)^2\right]+Y^2\cos^2\theta_P}\Big) \Big], \nonumber \\
\theta_T = 2\,&\textrm{atan}&\Big[\frac{1}{X\cos\theta_P\left[1-p(2\alpha-1)\right]} \times \Big(-Y\sin\theta_P  \nonumber \\
&\,& +\sqrt{X^2\cos^2\theta_P\left[1-p^2(2\alpha-1)^2\right]+Y^2\sin^2\theta_P}\Big) \Big], \nonumber
\end{eqnarray}
and $F^{1,1}$ is maximized for

$\left \{
\begin{array}{c @{\,=\,} c}
    X\sin\theta_P\left[\cos\theta_S-p(2\alpha-1)\right] +Y\cos\theta_P\sin\theta_S & 0 \\
    X\cos\theta_P\left[\cos\theta_T-p(2\alpha-1)\right] -Y\sin\theta_P\sin\theta_T & 0, \\
\end{array}
\right.$
which gives, for $\theta_P \neq 0,\pi$:
\begin{eqnarray}
\theta_S = 2\,&\textrm{atan}&\Big[\frac{1}{X\sin\theta_P\left[1+p(2\alpha-1)\right]} \times\Big(Y\cos\theta_P \nonumber \\
&\,& +\sqrt{X^2\sin^2\theta_P\left[1-p^2(2\alpha-1)^2\right]+Y^2\cos^2\theta_P}\Big) \Big], \nonumber \\
\theta_T = 2\,&\textrm{atan}&\Big[\frac{1}{X\cos\theta_P\left[1+p(2\alpha-1)\right]} \times \Big(-Y\sin\theta_P  \nonumber \\
&\,& +\sqrt{X^2\cos^2\theta_P\left[1-p^2(2\alpha-1)^2\right]+Y^2\sin^2\theta_P}\Big) \Big], \nonumber
\end{eqnarray}
with $\phi_S = \phi_T=\varphi$, $X = (2\eta-1)\sqrt{\alpha(1-\alpha)}$ and $Y = \left[p(2\alpha-1)^2-1\right]/2$.\\

\begin{figure*}[bht]
\centering
\includegraphics[width=1\textwidth]{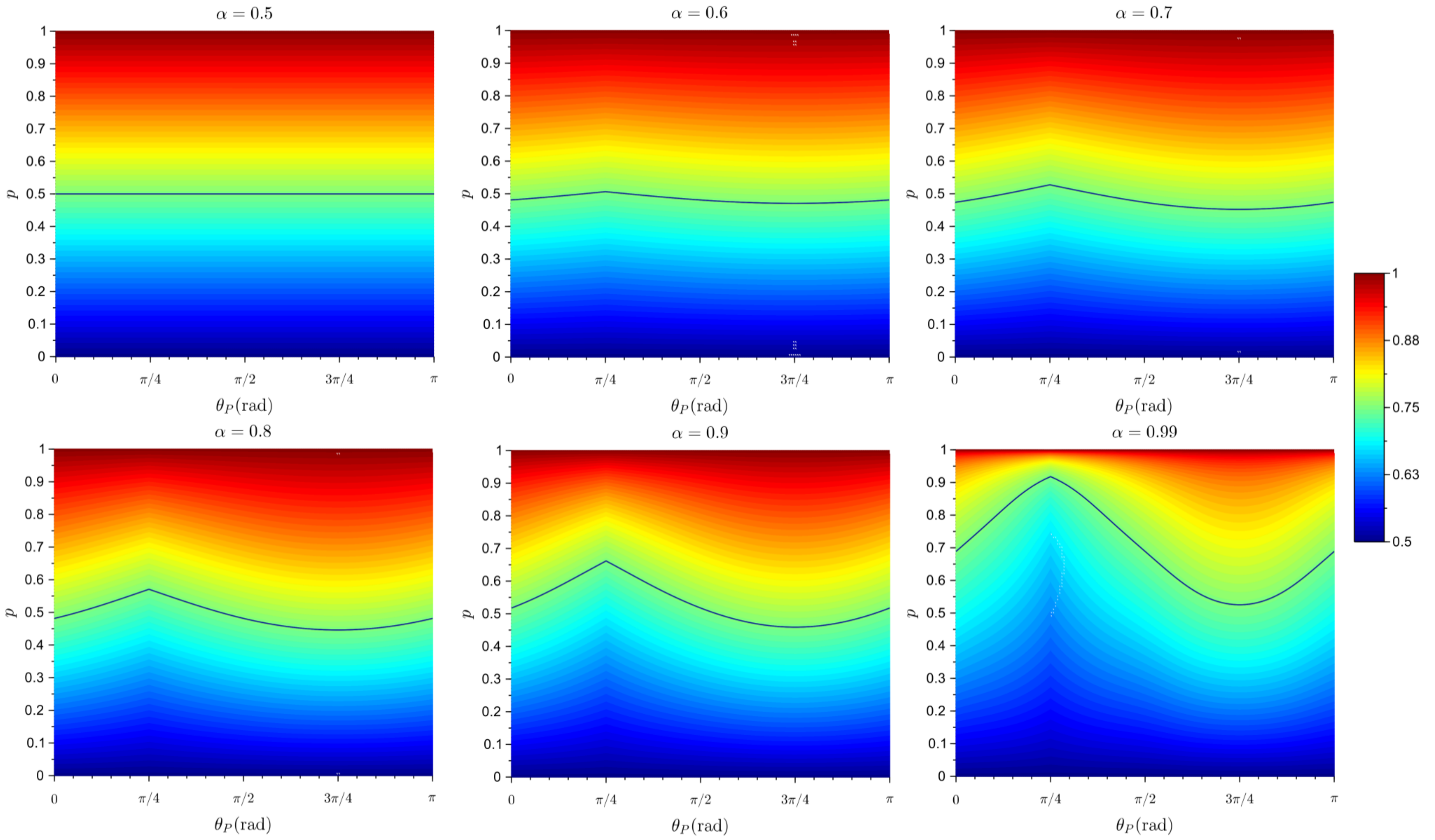}
\caption{Value of $F$ as a function of $\theta_P$ and $p$ for 6 different values of $\alpha$, with $\eta=1$. The black line in the green region on each plot corresponds to $F=3/4$.}
\label{Fig:FvsThetaPandp}
\end{figure*}

\medskip

\noindent We remark that
\begin{itemize}
\item $F=F^{0,0}$ for $\theta_P \in [0;\frac{\pi}{4}]$ and $\alpha \in [\frac{1}{2};1]$ or for $\theta_P \in [\frac{\pi}{4};\pi]$ and $\alpha \in [0;\frac{1}{2}]$,
\item $F=F^{1,1}$ for $\theta_P \in [0;\frac{\pi}{4}]$ and $\alpha \in [0;\frac{1}{2}]$ or for $\theta_P \in [\frac{\pi}{4};\pi]$ and $\alpha \in [\frac{1}{2};1]$,
\item $\cos(\theta_P+\frac{\pi}{4}) \geq 0$ for $\theta_P \in [0;\frac{\pi}{4}]$ and $\cos(\theta_P+\frac{\pi}{4}) < 0$ for $\theta_P \in ]\frac{\pi}{4};\pi]$,
\item $2\alpha-1 = |2\alpha-1|$ for $\alpha \in [\frac{1}{2};1]$ and $2\alpha-1 = -|2\alpha-1|$ for $\alpha \in [0;\frac{1}{2}]$,
\item $1-2\alpha = |2\alpha-1|$ for $\alpha \in [0;\frac{1}{2}]$ and $1-2\alpha = -|2\alpha-1|$ for $\alpha \in [\frac{1}{2};1]$.
\end{itemize}
Thus, with $\sigma=\textrm{sign}\big(\cos(\theta_P+\frac{\pi}{4})\big)$, we obtain the general expression of the fine-grained steering parameter (Eq.~(\ref{Eq:RhoSteeringPr})):
\begin{eqnarray}
F &=& \Big[ 1 + p\big[ \sigma|2\alpha-1|\cos\theta_P + \big(\sigma|2\alpha-1|+\cos\theta_P\big)\cos\theta_S \nonumber\\
&\,& + 2(2\eta-1)\sqrt{\alpha(1-\alpha)}\cos(\varphi-\phi_S)\sin\theta_P\sin\theta_S \big] \Big] \nonumber\\
&\,&\times \Big[4\big[1+p\sigma|2\alpha-1|\cos\theta_S\big]\Big]^{-1} \nonumber\\
&\times& \Big[ 1 + p\big[ -\sigma|2\alpha-1|\sin\theta_P + \big(\sigma|2\alpha-1|-\sin\theta_P\big)\cos\theta_T \nonumber\\
&\,& + 2(2\eta-1)\sqrt{\alpha(1-\alpha)}\cos(\varphi-\phi_T)\cos\theta_P\sin\theta_T \big] \Big] \nonumber\\
&\,&\times\Big[4\big[1+p\sigma|2\alpha-1|\cos\theta_T\big]\Big]^{-1}, \nonumber
\label{EqF}
\end{eqnarray}
for the optimal angles (for $\theta_P \neq 0,\pi$)
\begin{eqnarray}
\theta_S &=& 2\,\textrm{atan}\Big[\frac{1}{X\sin(\theta_P)\left[1-p\sigma|2\alpha-1|\right]} \times\Big(Y\cos(\theta_P) \nonumber \\
&\,& +\sqrt{X^2\sin^2(\theta_P)\left[1-p^2(2\alpha-1)^2\right]+Y^2\cos^2(\theta_P)}\Big) \Big], \nonumber \\
\label{EqFthetaSnon0} \\
\theta_T &=& 2\,\textrm{atan}\Big[\frac{1}{X\cos(\theta_P)\left[1-p\sigma|2\alpha-1|\right]} \times\Big(-Y\sin(\theta_P) \nonumber \\
&\,& +\sqrt{X^2\cos^2(\theta_P)\left[1-p^2(2\alpha-1)^2\right]+Y^2\sin^2(\theta_P)}\Big) \Big], \nonumber \\
\label{EqFthetaTnon0}
\end{eqnarray}
or (for $\theta_P = 0,\pi$)
\begin{eqnarray}
\theta_S &=& \theta_P, \label{EqFthetaS0} \\
\theta_T &=& \sigma\textrm{acos}\left(-p\sigma|2\alpha-1|\right).
\label{EqFthetaT0}
\end{eqnarray}
All the analytical expressions of optimal angles were calculated with \textit{Wolfram Alpha}~\cite{Wolfram}.

\section{Bob's choice of measurement settings and noise bounds for Scenario II}
\label{AppendixB}

In Fig.~\ref{Fig:FvsThetaPandp}, we plot the value of $F$ as a function of the measurement angle $\theta_P$ and the noise parameter $p$ for different values of $\alpha$, with $\eta=1$. For these plots, $\theta_S$ and $\theta_T$ have been optimized as above and $\theta_Q=\theta_P+\pi/2$. We see that for a given value of $p$, the minimum value of $F$ is obtained for $\theta_P=\pi/4$, except for $\alpha=1/2$ for which every choice of $\theta_P$ gives the same value of $F$ for a given value of $p$. The steering bound on $p$ as a function of $\alpha$ (for Scenario II) given in Fig.~\ref{Fig:pbounds} and \ref{Fig:BoundII} corresponds to the value of $p$ for which $F=3/4$ with $\theta_P=\pi/4$.

\section{Experimental probabilities and error bars}
\label{AppendixC}

\begin{figure*}[ht]
\centering
\includegraphics[width=1\textwidth]{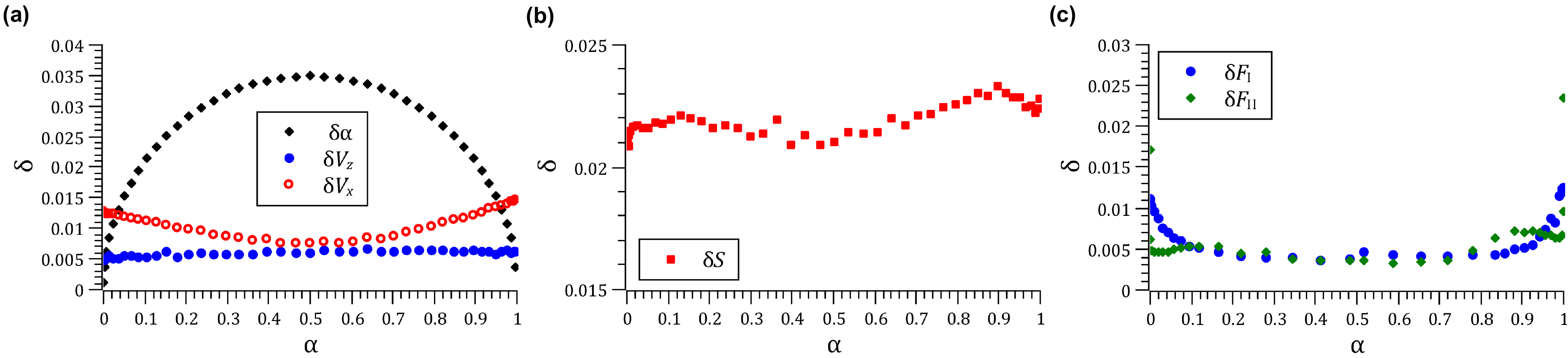}
\caption{(a) Setting errors on $\alpha$ and statistical errors on $V_z$ and $V_x$ ; (b) statistical errors on $S$ ; (c) statistical errors on $F_\textrm{I}$ and $F_\textrm{II}$, corresponding to the experimental results presented in Fig.~\ref{Fig:results}.}
\label{Fig:ErrorBars}
\end{figure*}

\subsection{Probabilities}

The joint probabilities $\textrm{P}(a_A,b_B)$ are experimentally estimated from coincidence counting of single-photon detection events in both avalanche photodiodes of the source presented in Fig.~\ref{Fig:Setup}. For each joint measurement setting $A \otimes B$, four such coincidence counts $C(\vartheta_A,\vartheta_B)$ are recorded with Alice's and Bob's polarizers set at the angles $\{\vartheta_A;\vartheta_B\}=\{\frac{\theta_A}{2};\frac{\theta_B}{2}\}$, $\{\frac{\theta_A}{2};\frac{\theta_B}{2}+\frac{\pi}{2}\}$, $\{\frac{\theta_A}{2}+\frac{\pi}{2};\frac{\theta_B}{2}\}$  and $\{\frac{\theta_A}{2}+\frac{\pi}{2};\frac{\theta_B}{2}+\frac{\pi}{2}\}$, corresponding to the measurement outcomes $\{0_A;0_B\}$, $\{0_A;1_B\}$, $\{1_A;0_B\}$ and $\{1_A;1_B\}$ respectively. Thus the experimental coincidence probabilities of obtaining the outcomes $a$ and $b$ for the measurement setting $A \otimes B$ are:
\begin{eqnarray}
\textrm{P}(0_A,0_B) &=& C\Big(\frac{\theta_A}{2},\frac{\theta_B}{2}\Big)\Big/C_{\textrm{tot}} \nonumber \\
\textrm{P}(0_A,1_B) &=& C\Big(\frac{\theta_A}{2},\frac{\theta_B}{2}+\frac{\pi}{2}\Big)\Big/C_{\textrm{tot}} \nonumber \\
\textrm{P}(1_A,0_B) &=& C\Big(\frac{\theta_A}{2}+\frac{\pi}{2},\frac{\theta_B}{2}\Big)\Big/C_{\textrm{tot}} \nonumber \\
\textrm{P}(1_A,1_B) &=& C\Big(\frac{\theta_A}{2}+\frac{\pi}{2},\frac{\theta_B}{2}+\frac{\pi}{2}\Big)\Big/C_{\textrm{tot}} \nonumber
\end{eqnarray}
with
\begin{eqnarray}
C_{\textrm{tot}} &=& C\Big(\frac{\theta_A}{2},\frac{\theta_B}{2}\Big)+C\Big(\frac{\theta_A}{2},\frac{\theta_B}{2}+\frac{\pi}{2}\Big) \nonumber \\
&\,& +C\Big(\frac{\theta_A}{2}+\frac{\pi}{2},\frac{\theta_B}{2}\Big)+C\Big(\frac{\theta_A}{2}+\frac{\pi}{2},\frac{\theta_B}{2}+\frac{\pi}{2}\Big) \nonumber
\end{eqnarray}
Thus, the visibilities are given by:
\begin{equation}
V_z = \frac{|C(0,0)+C(\frac{\pi}{2},\frac{\pi}{2})-C(0,\frac{\pi}{2})-C(\frac{\pi}{2},0)|}{C(0,0)+C(0,\frac{\pi}{2})+C(\frac{\pi}{2},0)+C(\frac{\pi}{2},\frac{\pi}{2})} \nonumber
\end{equation}
\begin{equation}
V_x = \frac{|C(\frac{\pi}{4},\frac{\pi}{4})+C(\frac{3\pi}{4},\frac{3\pi}{4})-C(\frac{\pi}{4},\frac{3\pi}{4})-C(\frac{3\pi}{4},\frac{\pi}{4})|}{C(\frac{\pi}{4},\frac{\pi}{4})+C(\frac{\pi}{4},\frac{3\pi}{4})+C(\frac{3\pi}{4},\frac{\pi}{4})+C(\frac{3\pi}{4},\frac{3\pi}{4})} \nonumber
\end{equation}
The CHSH parameter is given by:
\begin{equation}
S = |\textrm{E}(A \otimes B) + \textrm{E}(A \otimes B') + \textrm{E}(A' \otimes B) + \textrm{E}(A' \otimes B')| \nonumber
\end{equation}
with
\begin{eqnarray}
\textrm{E}(A \otimes B) &=& \frac{1}{C_{\textrm{tot}}} \Big[ C\Big(\frac{\theta_A}{2},\frac{\theta_B}{2}\Big)+C\Big(\frac{\theta_A}{2},\frac{\theta_B}{2}+\frac{\pi}{2}\Big) \nonumber \\
&\,& +C\Big(\frac{\theta_A}{2}+\frac{\pi}{2},\frac{\theta_B}{2}\Big)+C\Big(\frac{\theta_A}{2}+\frac{\pi}{2},\frac{\theta_B}{2}+\frac{\pi}{2}\Big)\Big] \nonumber
\end{eqnarray}
and $\theta_A=0$, $\theta_{A'}=\pi/2$ and $\theta_B$ and $\theta_{B'}$ given by Eq.~\ref{EqSthetaB}.\\
The fine-grained steering parameter is given by:
\begin{equation}
F = \displaystyle \max\left(F^{0,0},F^{1,1}\right) \nonumber
\end{equation}
with
\begin{eqnarray}
F^{0,0} &=& \frac{1}{2}C\Big(\frac{\theta_S}{2},\frac{\theta_P}{2}\Big)\Big/\Big[C\Big(\frac{\theta_S}{2},\frac{\theta_P}{2}\Big)+C\Big(\frac{\theta_S}{2},\frac{\theta_P}{2}+\frac{\pi}{2}\Big)\Big] \nonumber \\
&+& \frac{1}{2}C\Big(\frac{\theta_T}{2},\frac{\theta_Q}{2}\Big)\Big/\Big[C\Big(\frac{\theta_T}{2},\frac{\theta_Q}{2}\Big)+C\Big(\frac{\theta_T}{2},\frac{\theta_Q}{2}+\frac{\pi}{2}\Big)\Big] \nonumber
\end{eqnarray}
and
\begin{eqnarray}
F^{1,1} = \frac{1}{2}C\Big(\frac{\theta_S}{2}+\frac{\pi}{2},\frac{\theta_P}{2}+\frac{\pi}{2}\Big)&\Big/&\Big[C\Big(\frac{\theta_S}{2}+\frac{\pi}{2},\frac{\theta_P}{2}+\frac{\pi}{2}\Big) \nonumber \\
&\,& +C\Big(\frac{\theta_S}{2}+\frac{\pi}{2},\frac{\theta_P}{2}\Big)\Big] \nonumber \\
+ \frac{1}{2}C\Big(\frac{\theta_T}{2}+\frac{\pi}{2},\frac{\theta_Q}{2}+\frac{\pi}{2}\Big)&\Big/&\Big[C\Big(\frac{\theta_T}{2}+\frac{\pi}{2},\frac{\theta_Q}{2}+\frac{\pi}{2}\Big) \nonumber \\
&\,& +C\Big(\frac{\theta_T}{2}+\frac{\pi}{2},\frac{\theta_Q}{2}\Big)\Big] \nonumber
\end{eqnarray}
with $\theta_S$ given by Eq.~\ref{EqFthetaSnon0} or \ref{EqFthetaS0}, $\theta_T$ given by Eq.~\ref{EqFthetaTnon0} or \ref{EqFthetaT0}, $\theta_P=0$ for Scenario I or $\theta_P=\pi/4$ for Scenario II, and $\theta_Q=\pi/2$ for Scenario I or $\theta_Q=3\pi/4$ for Scenario II.

\subsection{Error bars}

The statistical errors on these probabilities and on $V_z$, $V_x$, $S$ and $F$ are estimated by propagating the Poissonian statistical error $\delta C$ on each coincidence count $C(\vartheta_A,\vartheta_B)$: $\delta C = \sqrt{C(\vartheta_A,\vartheta_B)}$. For each value of $\alpha$, we counted the coincidence events during 3 seconds to obtain each $C(\vartheta_A,\vartheta_B)$, thus obtaining around 5500 coincidence counts for the sum of the four possible joint outcomes $C_\textrm{tot}(\vartheta_A,\vartheta_B)=C(\vartheta_A,\vartheta_B)+C(\vartheta_A,\vartheta_B+\frac{\pi}{2})+C(\vartheta_A+\frac{\pi}{2},\vartheta_B)+C(\vartheta_A+\frac{\pi}{2},\vartheta_B+\frac{\pi}{2})$.
\begin{equation}
\delta V_z = 2\sqrt{\frac{\big[C(0,0)+C(\frac{\pi}{2},\frac{\pi}{2})\big]\big[C(0,\frac{\pi}{2})+C(\frac{\pi}{2},0)\big]}{C_\textrm{tot}^3(0,0)}} \nonumber
\end{equation}
\begin{equation}
\delta V_x = 2\sqrt{\frac{\big[C(\frac{\pi}{4},\frac{\pi}{4})+C(\frac{3\pi}{4},\frac{3\pi}{4})\big]\big[C(\frac{\pi}{4},\frac{3\pi}{4})+C(\frac{3\pi}{4},\frac{\pi}{4})\big]}{C_\textrm{tot}^3(\frac{\pi}{4},\frac{\pi}{4})}} \nonumber
\end{equation}
\begin{equation}
\delta S = \sqrt{\delta \textrm{E}(A \otimes B)^2 + \delta \textrm{E}(A \otimes B')^2 + \delta \textrm{E}(A' \otimes B)^2 + \delta \textrm{E}(A' \otimes B')^2 } \nonumber
\end{equation}
with
\begin{eqnarray}
\delta \textrm{E}(A \otimes B) &=& \frac{2}{\sqrt{C_\textrm{tot}^3(\frac{\theta_A}{2},\frac{\theta_B}{2})}}\sqrt{C\Big(\frac{\theta_A}{2},\frac{\theta_B}{2}\Big)+C\Big(\frac{\theta_A}{2}+\frac{\pi}{2},\frac{\theta_B}{2}+\frac{\pi}{2}\Big)} \nonumber \\
&\,& \times \sqrt{C\Big(\frac{\theta_A}{2},\frac{\theta_B}{2}+\frac{\pi}{2}\Big)+C\Big(\frac{\theta_A}{2}+\frac{\pi}{2},\frac{\theta_B}{2}\Big)} \nonumber
\end{eqnarray}
\begin{eqnarray}
\delta F^{0,0} &=& \frac{1}{2}\Bigg[\frac{C(\frac{\theta_S}{2},\frac{\theta_P}{2})C(\frac{\theta_S}{2},\frac{\theta_P}{2}+\frac{\pi}{2})}{\big[C(\frac{\theta_S}{2},\frac{\theta_P}{2})+C(\frac{\theta_S}{2},\frac{\theta_P}{2}+\frac{\pi}{2})\big]^3} \nonumber \\
&\,& +\frac{C(\frac{\theta_T}{2},\frac{\theta_Q}{2})C(\frac{\theta_T}{2},\frac{\theta_Q}{2}+\frac{\pi}{2})}{\big[C(\frac{\theta_T}{2},\frac{\theta_Q}{2})+C(\frac{\theta_T}{2},\frac{\theta_Q}{2}+\frac{\pi}{2})\big]^3}\Bigg]^{\frac{1}{2}} \nonumber
\end{eqnarray}
\begin{eqnarray}
\delta F^{1,1} &=& \frac{1}{2}\Bigg[\frac{C(\frac{\theta_S}{2}+\frac{\pi}{2},\frac{\theta_P}{2}+\frac{\pi}{2})C(\frac{\theta_S}{2}+\frac{\pi}{2},\frac{\theta_P}{2})}{\big[C(\frac{\theta_S}{2}+\frac{\pi}{2},\frac{\theta_P}{2}+\frac{\pi}{2})+C(\frac{\theta_S}{2}+\frac{\pi}{2},\frac{\theta_P}{2})\big]^3} \nonumber \\
&\,& +\frac{C(\frac{\theta_T}{2}+\frac{\pi}{2},\frac{\theta_Q}{2}+\frac{\pi}{2})C(\frac{\theta_T}{2}+\frac{\pi}{2},\frac{\theta_Q}{2})}{\big[C(\frac{\theta_T}{2}+\frac{\pi}{2},\frac{\theta_Q}{2}+\frac{\pi}{2})+C(\frac{\theta_T}{2}+\frac{\pi}{2},\frac{\theta_Q}{2})\big]^3}\Bigg]^{\frac{1}{2}} \nonumber
\end{eqnarray}
and $\delta F = \delta F^{0,0}$ or $\delta F = \delta F^{1,1}$.\\
The error on $\alpha$ is determined by the error $\delta \chi = \pm 1^\circ$ on the setting of the angle $\chi$ of the half-wave plate of the pump. As $\alpha=\cos^2(2\chi)$, the error on $\alpha$ is given by $\delta \alpha=2|\sin(4\chi)|\delta \chi$.\\
All experimental error bars are plotted in Fig.~\ref{Fig:ErrorBars}.

\bibliography{SteeringExp_bib}

\end{document}